\documentclass[prl, twocolumn, amsmath, amssymb, amsfonts]{revtex4}  % OR preprint

\usepackage{bm, graphics}
\pdfoutput=1
\usepackage[backref=false,colorlinks=true,breaklinks=true, citecolor=cyan]{hyperref}

\begin{document}

\title{Flux locking a superfluid interferometer}
\author{Yuki Sato}
\author{Aditya Joshi}
	\email{adityaj@berkeley.edu}
\author{Richard Packard}
\affiliation{Physics Department, University of California, Berkeley, California 94720, USA}

%ABSTRACT
\begin{abstract}
We demonstrate a flux locking technique using injected heat current to linearize the output of a superfluid helium-4 interferometer. A rotation flux through the interferometer loop produces a shift in the phase of the superfluid order parameter. This shift is nullified via negative feedback by a phase-shift caused by the injected heat current. The feedback signal is then a linear function of rotation flux. \\ \\
\begin{tabular}{|p{0.78 \textwidth}|}
\hline
Reproduced from \href{http://link.aip.org/link/?APL/91/074107}{Appl. Phys. Lett. 91, 074107 (2007)}, Copyright 2007, American Institute of Physics. This article may be downloaded for personal use only. Any other use requires prior permission of the author and the American Institute of Physics. 
\\
\hline
\end{tabular}
\end{abstract}

%Create the Title block
\maketitle

%START ACTUAL DOCUMENT

Interferometers are widely used in basic and applied science. These instruments, using sound, light, or de Broglie matter waves typically have a transfer function wherein the output amplitude (e.g. the Josephson critical current in a dc Superconducting Quantum Interference Device \cite{Clarke}(SQUID) ) is a cosinusoidally varying function of some variable of interest (magnetic flux in the case of the SQUID). To achieve widespread practical utility it is very useful to have some method to linearize the instrument's response. We report here a method by which this can be achieved for a superfluid 4He quantum interference device (SHeQUID), a superfluid analog of a superconducting dc SQUID. 

\begin{figure}[htbp]
\centering
	\includegraphics{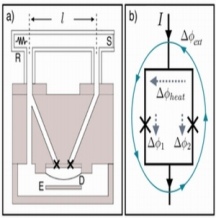}
	\caption{(Color online) a) Experimental apparatus. The inside is filled with superfluid $^4He$ and the entire apparatus is immersed in a bath of liquid helium. b) Equivalent SQUID circuit.  $\Delta\phi_{ext}$ is the phase-shift produced by some (possibly globally acting) external influence that the SHeQUID is being used to measure. $\Delta\phi_{1}$  and $\Delta\phi_{1}$  are the phase differences across the two weak-links and  $\Delta\phi_{heat}$ is the phase-shift due to injected heater power at R. }
	\label{fig1}
\end{figure}

The superfluid state in $^4He$ is described by a macroscopic order parameter written as $\psi = \left|\ psi \right|e^{i \phi}$  where  $\phi$ is the quantum phase. A SHeQUID [see Fig.~\ref{fig1}(a)] consists of two "weak-links" (marked X) placed in a loop filled with superfluid $^4He$ that is hydraulically coupled to a flexible diaphragm (D). A weak-link here consists of (nominally) 30nm diameter apertures spaced 3 $\mu m$ apart in a $100\times100$ square lattice on a 60nm thick silicon nitride membrane. The superfluid within these arrays oscillates at the Josephson frequency ($\omega_J = 2\pi f_J = \Delta\mu/\hbar$) when a chemical potential difference  $\Delta\mu$ is applied across them \cite{Hoskinson1}, \cite{Hoskinson2}. The chemical potential difference is applied by electrically pulling on the diaphragm (D) with the electrode (E). The diaphragm serves as the input element of a sensitive displacement sensor \cite{Chan}, which detects the oscillations. The heater (R) and sink (S) are used to inject a heat current into the top arm, thus producing a superfluid counterflow, which corresponds to a phase-drop ($\Delta\phi_{heat}$) between the ends of the tube \cite{Sato}.

Depending on the temperature, the weak links oscillate either as $\sin \phi$ Josephson weak links or coherent phase slip centers \cite{Hoskinson2}, \cite{Sukhatme}.  In either case, each of the weak links emits a strong Fourier component at frequency $f_J$. Let the amplitudes of these first harmonics in the two arrays be represented as  $I_{0,1}$ and $I_{0,2}$. The superposition of the two oscillations detected by the microphone can be written as $I_{total} = I_t \sin \left[\omega_J t + \delta \right]$,  where the interference amplitude is given by:

\begin{equation}\label{eq1}
	I_t = I_0 \left[ \cos^2 \theta + \gamma^2 \sin^2 \theta \right]^{1/2} \equiv I_0 F_{\gamma}(\theta)
\end{equation}

Here $I_0 \equiv I_{0,1} + I_{0,2}$, $\theta \equiv \left|\Delta \phi_1 - \Delta \phi_2\right|/2$   is half the difference in the phase-drops of the two oscillators, and $F_{\gamma}(\theta)$  is a dimensionless non-linear periodic function in $\theta$ with an asymmetry parameter defined as $\gamma =\equiv \left(I_{0,1} - I_{0,2}\right)/\left(I_{0,1} + I_{0,2}\right)$.

Single-valuedness of the order parameter demands that  $\oint \vec{\nabla}\phi \cdot \vec{dl} = 2\pi n$ for integer n (where the phase integral goes around the interferometer loop). When no currents flow \cite{note1} in the interferometer, there are no phase gradients and $\oint \vec{\nabla}\phi \cdot \vec{dl} = 0$. This phase integral condition is maintained even if an external influence induces flow in the sense loop as long as flow velocities remain sufficiently low (i.e. below the velocity to create quantum vortices so that n remains 0). If  $\Delta \phi_{ext}$ is the shift in the phase of the macroscopic wave function created by some external influence and  $\Delta \phi_{heat}$ the phase-shift due to a heat current in the top arm, the circulation quantization condition allows us to write  $\Delta \phi_{ext} + \Delta \phi_{heat} + \Delta \phi_{1} - \Delta \phi_{2} = 0$  [see Fig.~\ref{fig1}(b)]. The phase differences across the remaining segments of the loop are made negligible (by design). Then Eq.~\ref{eq1} becomes:
 			
\begin{equation}\label{eq2}
	I_t = I_0 F_{\gamma}\left( \frac{\Delta \phi_{ext}}{2} + \frac{\Delta \phi_{heat}}{2}  \right)
\end{equation} 			
 			
\begin{figure}[htbp]
		\includegraphics{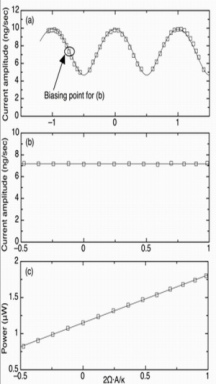}
	\caption{a) Current amplitude (ng/s) modulation due to changes in the Earth's rotation flux (via reorientation of loop area) in the absence of heater power biasing. b) Modulation compensated by injected heater current thus making the amplitude independent of the rotation flux. Amplitude is maintained constant at the bias point circled in (a). c) Feedback heater power ($\mu W$) needed for a given value of rotation flux to maintain the oscillation amplitude at a constant level as shown in (b). This heater power serves as a linear measure of the rotation that it is used to nullify. This effectively gives us a flux-locked SHeQUID. These data are taken at $T_{\lambda} - T \approx 16 mK$.}
	\label{fig2}
\end{figure}

For example, in previous work \cite{Hoskinson3} the external phase shifts were produced by the (steady) rotation field of the Earth, which creates a rotation flux  $\vec{\Omega} \cdot \vec{A}$ in the SHeQUID ( $\vec{\Omega}$ is the angular velocity vector of the Earth and $\vec{A}$ is the area vector of the interferometer loop). This rotation flux induces a so-called Sagnac phase-shift \cite{Page}, \cite{Werner}  given by 
  		
\begin{equation}\label{eq3}
	\Delta \phi_{Sagnac} = (4\pi m_4 /h) \vec{\Omega} \cdot \vec{A}
\end{equation} 		  		
  		
where $m_4$ is the atomic mass of $^4He$. Using this in Eq.~\ref{eq2} (with no heat current), the SHeQUID current amplitude $I_t$  is proportional to  $F_{\gamma}\left( 2\pi \vec{\Omega} \cdot \vec{A} /\kappa \right)$ where  $\kappa$ is the $^4He$ quantum of circulation. Fig.~\ref{fig2}(a) shows the non-linear and periodic nature of the function $F_{\gamma}$  in Eq.~\ref{eq2}.

This non-linearity is problematic because the sensitivity of the SHeQUID (slope of the curve) varies as the rotation flux is changed (see Ref.\cite{Hoskinson3} for a discussion of this 'open loop gain' for the SHeQUID as a rotation sensor). Although this sensitivity is very high in the regions of steepest slope (e.g. the circled point in the figure), it falls off dramatically at the extrema. Also, because of the periodicity of the amplitude, large changes in rotation flux can be determined only by tracing the whole pattern (temporally impractical). To take advantage of the maximum sensitivity of the SHeQUID, it is necessary to bias the device at the point of greatest slope. Further, to develop a practical rotation sensor (i.e. gyroscope) it is important to be able to linearize the device to have an output directly proportional to rotation flux.

Following the example of the SQUID [Refs. \cite{Forgacs1} and \cite{Forgacs2}],  where the magnetic field of interest is canceled by a (negative) feedback magnetic field, one could imagine holding the current amplitude of the SHeQUID constant by an applied retrograde rotation of the loop to cancel the rotation of interest and thereby measure the original rotation stimulus. This is clearly an awkward mechanical solution.  Rather, we seek a more convenient phase shifting influence, which can be easily applied to the interferometer loop in order to implement this negative feedback scheme.

Fig.~\ref{fig1}(a) displays our solution to the problem. We show an interferometer loop in which one arm is a straight tube [of interior length $l=2.5 \pm 0.05 cm$  and circular cross-sectional area $\sigma = (3.78\pm 0.04)\times 10^{-2} cm^2$ ] containing a heater (R) at one end. The other end of the tube terminates in a roughened copper disk heat exchanger (S) that is the dominant thermal path connecting the fluid in the interferometer with the surrounding temperature-stabilized bath. This tube is made of Stycast 1266 (insulating) to minimize the heat loss through the walls. When power   is applied to the heater, the phase difference created across the tube's ends because of the steady superfluid counterflow set up in the tube is given by \cite{Sato}:

\begin{equation}\label{eq4}
	\Delta \phi_{heat} = \frac{l}{\sigma}\frac{2\pi m_4}{h} \frac{\rho_n}{\rho \rho_s T s} \dot{Q}
\end{equation}

where $\rho$, $\rho_n$ and $\rho_s$ are the total, normal and superfluid densities respectively, s is the entropy per unit mass and T is the temperature in the cell.

As before,  $\Delta \phi_{ext}$ in Fig.~\ref{fig1}(b) is the Sagnac phase-shift due to the Earth. Equation~\ref{eq2} combined with Eqs.~\ref{eq3} and ~\ref{eq4} for the phase-shifts then becomes: 

\begin{equation}\label{eq5} 	
I_t = I_0 F_{\gamma}\left( a\vec{\Omega} \cdot \vec{A} +  b \dot{Q} \right) 
\end{equation}

where  $a = \frac{2\pi m_4}{h}$ and  $b = \frac{l}{\sigma}\frac{\pi m_4}{h} \frac{\rho_n}{\rho \rho_s T s}$ are constants for a given temperature.

Any change in rotation flux can now be cancelled by injecting heater power to keep the argument of $F_{\gamma}$ constant in Eq.~\ref{eq5}. The interferometer can thus be maintained at fixed current amplitude and the flux is ``locked''. Further, the amount of power needed for this purpose provides a linear measure of the change in rotation flux: $\left| \vec{\Omega} \cdot \vec{A} \right| b \dot{Q}/a$ .

Fig.~\ref{fig2} demonstrates the operation of the feedback. Fig.~\ref{fig2}(a) shows the signature sinusoidal interference pattern due to the reorientation of the SHeQUID loop about the vertical with no feedback applied. This is the previously mentioned Sagnac effect caused by the Earth's rotation. The vertical axis is proportional to the measured amplitude of Josephson oscillations in the SHeQUID. We vary the rotation flux $\vec{\Omega} \cdot \vec{A}$  by changing the angle between the loop and the Earth's spin axis.

Fig.~\ref{fig2}(b) shows the same measured amplitude as in Fig.~\ref{fig2}(a), this time with power applied to the heater thereby creating a phase change in the heater tube, which compensates for the rotation flux change. Within the noise level of the experiment the SHeQUID current amplitude is now independent of rotation flux (i.e. $a \vec{\Omega} \cdot \vec{A} + b \dot{Q} = constant$ ).

Fig.~\ref{fig2}(c) shows the heater power $\dot{Q}$  that is injected to maintain the current amplitude constant plotted against rotation flux  $\vec{\Omega} \cdot \vec{A}$. Within the noise level it is seen that $\dot(Q) \propto \vec{\Omega} \cdot \vec{A}$ . The loop is now phase-locked and the output is linearized. Using the calibration obtained from Fig.~\ref{fig2}(c), incident rotation flux may be measured via this negative feedback mechanism. Similarly, using Eqs.~\ref{eq2} and ~\ref{eq4}, any unknown phase-shifting influence  $\Delta\phi_{ext}$ can be directly measured simply by monitoring the feedback heater power. In practice, this feedback scheme works for injected heater power values lower than that corresponding to 250 complete cycles in Fig.~\ref{fig2}(a) (i.e. for phase-shifts up to $250 \times 2\pi$ ). For heater power values greater than this limit (which varies with temperature but is about a few hundred microwatts), we observe a rapid onset of quantum turbulence, which renders the interferometer useless for measuring external influences. Noise spectra and drift considerations are discussed elsewhere \cite{Mukharsky}, \cite{AV}.

In conclusion, the flux locking method described here linearizes a SHeQUID so that this class of instrument can be used to monitor widely varying phase shifting influences such as rotation.

%ACKNOWLEDGEMENTS
\begin{acknowledgments}
We thank John Treichler for help with the fabrication of aperture arrays and E. Hoskinson for the early versions of the lab software as well as for sharing his insights on weak-link physics with one of us (Y.S). I. Hahn generously provided the high-resolution thermometer.  This work was supported in part by NSF Grant No. DMR 0244882 and by the ONR. The aperture arrays were fabricated at the Cornell NanoScale Facility, a member of the NSF National Nanotechnology Infrastructure Network. 
\end{acknowledgments}

\bibliography{fluxlockedSHeQUID}

\end{document}